\documentclass[sigconf]{acmart}
\acmConference[EASE 2026]{The 30th International Conference on Evaluation and Assessment in Software Engineering}{Tue 9 - Fri 12 June 2026}{Glasgow, United Kingdom}

\usepackage{amsmath}
\usepackage{graphicx}
\usepackage{textcomp}
\usepackage{graphicx}
\usepackage{array}
\usepackage{tcolorbox}
\usepackage{booktabs}
\usepackage{placeins}
\usepackage{listings} 
\usepackage{xcolor}   
\usepackage{hyperref}
\usepackage{algorithm}
\usepackage{algpseudocode}
\usepackage{multirow}
\usepackage{makecell}
\usepackage{colortbl} 
\usepackage{tabularx}
\usepackage{everypage}

\usepackage{booktabs}
\usepackage{pifont}

\usepackage{tikz}
\usetikzlibrary{
  arrows.meta,
  positioning,
  fit,
  backgrounds,
  calc,
  shapes.geometric
}

\newcommand{\cmark}{\ding{51}}%

\AtBeginDocument{%
  }

\setcopyright{acmlicensed}
\copyrightyear{2026}
\acmYear{2026}
\setcopyright{cc}
\setcctype{by}
\acmConference[EASE '26]{International Conference on Evaluation and Assessment in Software Engineering}{June 09--12, 2026}{Glasgow, United Kingdom}
\acmBooktitle{International Conference on Evaluation and Assessment in Software Engineering (EASE '26), June 09--12, 2026, Glasgow, United Kingdom}
\acmDOI{10.1145/3816483.3816557}
\acmISBN{979-8-4007-2348-3/2026/06}

\begin{document}

\title[From TinyGo to gc Compiler: Extending Zorya's Concolic Framework to Real-World Go Binaries]{From TinyGo to gc Compiler: Extending Zorya's\\ Concolic Framework to Real-World Go Binaries}


\author{Karolina Gorna}
  \affiliation{%
  \institution{Telecom Paris and Ledger Donjon}
  \city{Paris}
  \country{France}
 }
 \email{karolina.gorna@telecom-paris.fr}

 \author{Nicolas Iooss}
  \affiliation{%
  \institution{Ledger Donjon}
  \city{Zurich}
  \country{Switzerland}
  }
  \email{nicolas.iooss@ledger.com}

 \author{Yannick Seurin}
  \affiliation{%
  \institution{Ledger Donjon}
  \city{Paris}
  \country{France}
  }
  \email{yannick.seurin@ledger.com}

 \author{Rida Khatoun}
  \affiliation{%
  \institution{Telecom Paris}
  \city{Palaiseau}
  \country{France}
  }
  \email{rida.khatoun@telecom-paris.fr}

  \author{Keith Makan}
  \affiliation{%
  \institution{University of the Western Cape}
  \city{Cape Town}
  \country{South Africa}
  }
  \email{keith.makan@kminfosec.co.za}

\renewcommand{\shortauthors}{Gorna \textit{et al.}}

\begin{abstract}
Zorya is a concolic execution framework that lifts compiled binaries to Ghidra's P-Code intermediate representation and uses the Z3 SMT solver to detect vulnerabilities by reasoning over both concrete and symbolic values. Previous versions supported only single-threaded TinyGo binaries. In this paper, we extend Zorya to multi-threaded binaries produced by Go's standard \texttt{gc} compiler. This is achieved by restoring OS thread states from \texttt{gdb} dumps, neutralizing runtime preemption, and introducing overlay path analysis with copy-on-write semantics to detect silent vulnerabilities on untaken branches. We rigorously assess Zorya on 11 real-world vulnerabilities from production Go projects such as Kubernetes, Go-Ethereum, and CoreDNS. Our evaluation shows that Zorya detects seven bugs at the binary level, including a silent integer overflow detects no other evaluated tool finds without a manually written oracle.
\end{abstract}




\maketitle

\section{Introduction}
\label{intro}

The Go programming language is widely adopted in cloud infrastructure, container orchestration (e.g., Kubernetes~\cite{carrion_kubernetes_2022}), and blockchain systems \cite{staff_octoverse_2024}. Security analysis of Go binaries remains difficult because the standard \texttt{gc} compiler produces executables with a complex runtime. This runtime manages goroutines, garbage collection, and preemptive scheduling through multiple OS threads. Existing symbolic execution tools such as KLEE~\cite{zhang_survey_2020}, angr~\cite{wang_angr_2017}, and Radius2~\cite{aemmitt-ns_aemmitt-nsradius2_2025} were designed for C/C++ and do not model Go-specific mechanisms. Furthermore, most of them use LLVM as an intermediate representation (IR), to which Go doesn't compile fully.

Zorya was introduced by Gorna \textit{et al.}~\cite{gorna_concolic_2026,gorna_zorya_2025} as a concolic execution framework, that combines concrete and symbolic execution for code exploration and vulnerability detection. It lifts binaries to Ghidra's P-Code intermediate representation~\cite{nsa_ghidra_2017} and targets Go binaries compiled with TinyGo~\cite{tinygo-org_tinygo_2019-1}, a compiler that produces single-threaded executables with a minimal runtime. The second version added a panic-reachability gating filter to focus symbolic reasoning on panic-reachable paths.

However, TinyGo is rarely used in production. Real-world Go projects are compiled with \texttt{gc}~\cite{go-community_introduction_nodate}, the standard Go toolchain. A \texttt{gc} binary differs from a TinyGo binary in several fundamental ways. First, the \texttt{gc} runtime spawns multiple OS threads at startup to host goroutine scheduling. Second, the runtime uses cooperative and asynchronous preemption: a sentinel value in the goroutine descriptor forces function prologues to yield control. Third, the binary invokes Linux VDSO functions for time queries, which are memory-mapped at runtime and absent from the ELF file. These differences prevented the previous versions of Zorya from analyzing \texttt{gc} binaries. This paper presents the extensions required to bridge this gap. Specifically, our contributions are as follows:

\begin{itemize}
    \item \textbf{Concolic analysis of \texttt{gc}-compiled binaries}: This work extends Zorya to analyze multi-threaded binaries produced by the standard Go compiler.\footnote{\url{https://github.com/Ledger-Donjon/zorya}}
    \item \textbf{Real-world Go vulnerability dataset}: A corpus of 11 real-world Go vulnerabilities from production projects, with reproduction workflows and triggering inputs.\footnote{\url{https://github.com/Ledger-Donjon/logic_bombs_go}}
    \item \textbf{Comparative evaluation}: Zorya is evaluated against seven state-of-the-art tools (three static analyzers, two fuzzers, two binary-level symbolic executors) and detects 7/11 bugs, including a silent integer overflow no other tool finds without an oracle.\footnote{\url{https://github.com/Ledger-Donjon/zorya-evaluation}}
\end{itemize}

\textbf{Assumptions.} Zorya assumes correct Ghidra disassembly; execution halts if jumps target unidentified code. We mitigate this via preprocessing and compiler predictable layouts. Currently, Zorya analyzes non-interactive binaries requiring inputs at initialization.

\section{Background}

\subsection{Symbolic Execution and its Limits in Go}
Given a program $P$ with input variables $X = \{x_1, \ldots, x_n\}$, symbolic execution replaces each $x_i$ with a symbolic value $s_i$ and treats ope\-rations symbolically. At each branch on a predicate $\varphi(s_1, \ldots, s_n)$, the engine forks: one path adds $\varphi$ to the path condition $\Pi$, while the other adds $\neg\varphi$. An SMT solver~\cite{de_moura_z3_2008} then checks if $\Pi$ is solvable to see if a path is possible. To reduce \emph{path explosion}—where the number of paths grows too fast for the solver—\emph{concolic execution} performs both concrete and symbolic execution. It uses real values to drive the program's flow while keeping symbolic expressions to check other paths.

The Go runtime contains roughly 300,000 lines of code~\cite{go-community_gosrcruntimeracego_2026} and creates major hurdles for symbolic tools. Its complex, version-specific structures for goroutines, garbage collection, and stack management often trigger path explosion before the tool even reaches the user's code. Additionally, many engines do not model the low-level OS calls (like \texttt{futex} or \texttt{clone}) that Go needs. As a result, while tools like BINSEC~\cite{djoudi_binsec_2015} and SymQEMU~\cite{poeplau_symqemu_2021} offer some binary-level support, other frameworks like KLEE, angr and Radius2 often fail to run \texttt{gc}-compiled binaries at all~\cite{gorna_concolic_2026}.

\subsection{Go Analysis Tools}
The Go ecosystem provides several analysis tools. We briefly des\-cribe the most used ones and those used in our evaluation.

\textbf{Static analyzers.} \texttt{go~vet}~\cite{google_govet_2024} is the standard static checker shipped with the Go toolchain; it detects common mistakes such as unreachable code and incorrect format strings. \texttt{staticcheck}~\cite{google_staticcheck_2018} extends \texttt{go~vet} with additional style and correctness checks. Then, \texttt{gosec}~\cite{securego_gosec_2024} focuses on security patterns; its G115 rule flags type-conversion overflows (e.g., \texttt{uint64} to ~\texttt{int}) but does not check same-type arithmetic overflows. \texttt{govulncheck}~\cite{go-community_govulncheck_2026} queries the Go advisory database and reports known CVEs affecting the binary's dependencies. \texttt{nilaway}~\cite{mahajan_nilaway_2023} performs inter-procedural nil-flow inference to detect potential nil-pointer dereferences.

\textbf{Fuzzers.} \texttt{go~test~-fuzz}~\cite{go-community_go_2026} is the built-in coverage-guided fuzzer introduced in Go~1.18. It replaced the earlier \texttt{go-fuzz} tool~\cite{vyukov_dvyukovgo-fuzz_2024}, which is now deprecated. GoLibAFL~\cite{security-research-labs_srlabsgolibafl_2026} is a coverage-guided fuzzer based on LibAFL; it instruments Go binaries at the source level and achieves higher throughput than the built-in fuzzer. Both fuzzers require a manually written harness for each tested function.

\textbf{Support tools.} Go also ships with debugging and inspection utilities. Delve~\cite{go-community_go-delvedelve_2026} is the standard Go debugger, aware of goroutines and Go types. \texttt{go~tool~nm}~\cite{go-community_nm_2026} lists the symbols in a compiled binary, which is useful for locating functions and understanding binary layout. These tools are used alongside the analyzers above.

\subsection{The Zorya Concolic Framework}
Zorya is a concolic execution engine for analyzing compiled binaries. It lifts x86-64 binaries to Ghidra's P-Code, a register-transfer intermediate representation with approximately 70 opcodes. Each P-Code instruction is executed with both a concrete value (from a \texttt{gdb} memory dump) and a symbolic expression (maintained by the Z3 SMT solver). The concrete value drives control flow; the symbolic expression accumulates constraints.

In its initial version, Zorya supported only TinyGo-compiled Go binaries. TinyGo produces single-threaded executables with a cooperative scheduler and no preemption. The runtime is small, and most system interactions are direct syscalls. This simplified environment allowed Zorya to execute binaries end-to-end without modeling thread management or complex runtime internals.

The second version introduced a \emph{panic-reachability gating} filter that uses a precomputed reverse call graph to skip branches that cannot reach a panic site, yielding 1.8--3.9$\times$ speedups. 

\begin{figure*}[htbp]
\centerline{\includegraphics[scale=0.22]{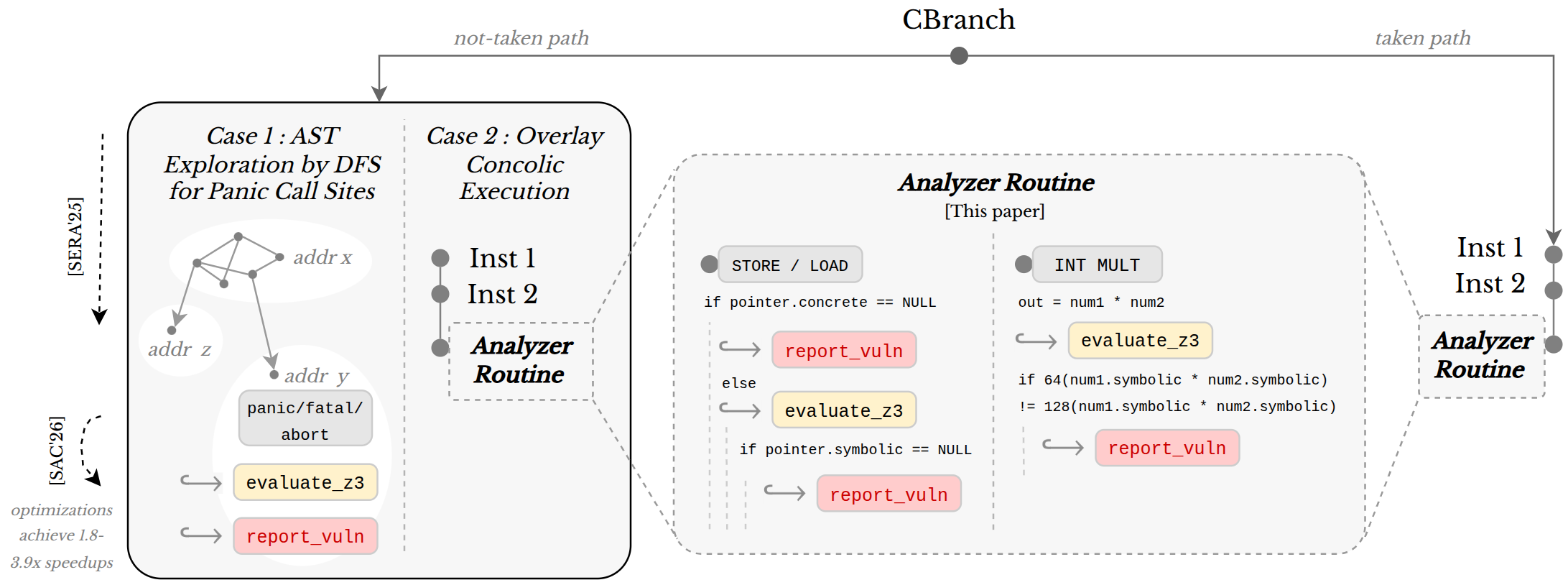}}
\caption{Overview of Zorya workflow, including AST exploration and Overlay Concolic Execution.}
\label{fig:zorya-overview}
 \begin{minipage}{\textwidth}
   \small\textit{Note: At each conditional branch (CBranch), the path not taken concretely is analyzed. For Go binaries, the AST is first explored from the corresponding node to locate \texttt{panic}, \texttt{fatal}, or \texttt{abort} call sites. If one is found, the SMT solver checks whether satisfying variable assignments exist (function or binary arguments, depending on the mode), indicating a potential vulnerability. Then, for all binary types, an overlay concolic execution is performed over the next 15 instructions to detect vulnerabilities that may occur without triggering a runtime panic. This \textit{Analyzer Routine} is applied to both the taken and not-taken paths; if the check succeeds, the solver reports the feasibility of the faulty path. The state is then restored and concolic execution resumes.}
   \end{minipage}
\label{fig}
\end{figure*}

In this paper, we further introduce \emph{overlay path analysis}, a copy-on-write mechanism~\cite{tran_survey_2024} that executes the untaken branch of a conditional to detect vulnerabilities such as null-pointer dereferences and integer overflows, as illustrated in Figure~\ref{fig:zorya-overview}.

\section{Related Work}
\label{sec:related}

\textbf{Selective path exploration.}
Classical symbolic execution forks the full state at every branch~\cite{baldoni_survey_2018}. SAGE~\cite{godefroid_sage_2012} negates collected constraints offline without forking; S2E~\cite{chipounov_s2e_2011} uses copy-on-write snapshots to reduce fork cost; Driller~\cite{stephens_driller_2016} invokes concolic execution only when a fuzzer stalls. Zorya's overlay shares S2E's copy-on-write principle but at a finer granularity: it explores the untaken branch for a bounded number of instructions, then discards the overlay. Our evaluation uses BINSEC and SymQEMU instead, as they operate directly on Linux ELF binaries—SAGE targets Windows PE, S2E requires a full VM image, and Driller depends on AFL instrumentation.

\textbf{Multi-threaded symbolic execution.}
Cloud9~\cite{bucur_parallel_2011} extends KLEE to model \texttt{pthread} primitives and explore thread interleavings for concurrency-bug detection. Zorya addresses a different problem: the Go runtime spawns OS threads before user code runs, and the analysis must manage them—restoring state from dumps and scheduling execution—so that the concolic engine can reach user-level functions.

\textbf{Integer overflow detection.}
IntScope~\cite{wang_intscope_nodate} detects overflows in x86 binaries via symbolic execution; IOC~\cite{dietz_understanding_2015} instruments C/C++ source to catch undefined integer behavior at runtime. Both target C/C++, where overflows are undefined. In Go, unsigned arithmetic wraps silently by specification. The closest contemporary effort, go\nobreakdash-panikint~\cite{valerio_detect_2025}, modifies the Go compiler to turn silent overflows into explicit panics for fuzzers.

\textbf{Positioning.} Prior work addresses each of these aspects in isolation; Zorya combines fine\nobreakdash-grained overlay exploration, multi\nobreakdash-thread initial\nobreakdash-state recovery, and silent\nobreakdash-overflow detection for Go binaries.

\section{Overlay Concolic Execution}
\label{sec:overlay}

\subsection{Copy-on-Write State}
A key challenge in concolic execution is reasoning about paths not taken by the concrete trace. Classical approaches fork the entire state at each branch, leading to exponential growth~\cite{baldoni_survey_2018}. Zorya instead uses an \emph{overlay} mechanism inspired by copy-on-write semantics: when a conditional branch involves tracked symbolic variables, Zorya creates a lightweight overlay for the untaken path. Reads first consult the overlay and fall through to the base state on a miss; writes go exclusively to the overlay. This avoids cloning the full execution state, which can exceed 1\,GiB for \texttt{gc} binaries with large heaps, so only registers and memory bytes actually modified during overlay execution incur a copy.

\subsection{Overlay Execution Protocol}
The overlay execution proceeds as follows:

\begin{enumerate}
\item Save the executor's mutable state: unique temporary variables, current address, call stack depth, freed stack frames, and the null-check cache.
\item Set the overlay's instruction pointer (RIP) to the untaken branch target.
\item Execute up to $N$ P-Code instruction blocks (default $N\!=\!15$) using the standard executor, which transparently reads from and writes to the overlay.
\item At each instruction, check for vulnerability patterns \emph{before} execution:
null-pointer dereferences (loads and stores) and integer overflows; and, not yet
evaluated, division by zero and accesses to freed stack frames.
\item On finding a vulnerability, record it and return. On reaching a \texttt{RETURN}, a loop, or the depth limit, stop.
\item Discard the overlay. Restore all saved state so the main execution path is unaffected.
\end{enumerate}

The null-check cache deserves special attention. During overlay execution, SAT results (confirmed nullable pointers) are retained because the vulnerability is real regardless of the path. UNSAT results (proven non-null under the real path's constraints) are cleared because the negated branch may make previously-safe pointers nullable.

\subsection{Illustrative Example: Overlay Detection of an Integer Overflow}
\label{sec:example}

We illustrate Zorya's integer overflow detector on Go-Ethereum's \texttt{memoryGasCost} function (geth~v1.6.0), shown in Listing~\ref{lst:gasCost}. The guard on line~3 prevents \texttt{toWordSize} from overflowing, but does \emph{not} prevent the squaring on line~7 from wrapping: for large inputs the true product exceeds 64~bits and Go silently truncates it, letting an attacker allocate massive EVM memory while paying near-zero gas. Zorya runs this function with concrete arguments that follow a valid, non-overflowing path. The overflow is detected \emph{entirely through the symbolic expressions} attached to each variable. Ghidra lifts the multiplication to a P-Code \texttt{INT\_MULT} instruction.

\begin{lstlisting}[language=Go, caption={Vulnerable \texttt{memoryGasCost} in Go-Ethereum v1.6.0 . }, label={lst:gasCost}]
func memoryGasCost(mem *Memory,
        newMemSize uint64) (uint64, error) {
  if newMemSize > MaxUint64-32 {
    return 0, errGasUintOverflow
  }
  newMemSizeWords := toWordSize(newMemSize)
  square := newMemSizeWords * newMemSizeWords
  linCoef := newMemSizeWords * MemoryGas
  quadCoef := square / QuadCoeffDiv
  fee := linCoef + quadCoef - mem.lastGasCost
  return fee, nil
}
\end{lstlisting}

\begin{table*}[t]
\centering
\caption{Evaluation of Go toolchain tools, symbolic execution tools and Zorya against a real-world Go bugs dataset. \cmark~indicates caught vulnerability; empty cells indicate the vulnerability was not caught.}
\label{tab:go-tools-comparison}
\footnotesize
\renewcommand{\arraystretch}{1.2}
\resizebox{\textwidth}{!}{%
\begin{tabular}{@{} l l c *{5}{c} | c >{\centering\arraybackslash}p{2.2cm} | c c c @{}}
\toprule
& & \textbf{Size (MB)} & \textbf{staticcheck} & \textbf{gosec} & \textbf{nilaway} & \textbf{go-fuzz} & \textbf{GoLibAFL}
& \multicolumn{2}{c|}{\textbf{Zorya}}
& \textbf{BINSEC} & \textbf{SymQEMU} & \textbf{KLEE} \\
\midrule
\multicolumn{2}{@{}l}{\textit{Technique used}}
& & \shortstack{Static\\Analysis} & \shortstack{Static\\Analysis} & \shortstack{Static\\Analysis} & Fuzzing & Fuzzing
& \multicolumn{2}{c|}{Concolic Execution}
& \shortstack{Symbolic\\Execution} & \shortstack{Symbolic\\Execution} & \shortstack{Symbolic\\Execution} \\
\midrule
\multicolumn{2}{@{}l}{\shortstack[l]{\textit{Needs Additional Files or}\\\textit{Automated Execution (Auto.)\,?}}}
& & Auto. & Auto. & Auto. & \textbf{Harness} & \textbf{Harness}
& \multicolumn{2}{c|}{Auto.}
& \textbf{Init Files} & Auto. & \textbf{LLVM bitcode} \\
\midrule
\multicolumn{2}{@{}l}{\shortstack[l]{\textit{Outputs the execution trace}\\\textit{of each instruction\,?}}}
& & No & No & No & No & No
& \multicolumn{2}{c|}{\textbf{Yes}}
& \textbf{Yes} & No & \textbf{Yes} \\
\midrule
\multirow{4}{*}{\shortstack[l]{\textit{Nil Pointer}\\\textit{Dereference}}}
& \mbox{kubectl-2025}      & 106 & . & . & . & \cmark & \cmark & \cmark & \makecell[c]{Analyzer routine\\(LOAD check)} & . & . & . \\
\arrayrulecolor{gray!40}\cline{2-13}\arrayrulecolor{black}
& \mbox{kubelet-2025}      & 121 & . & . & . & \cmark & \cmark & \cmark & \makecell[c]{Analyzer routine\\(LOAD check)} & . & . & . \\
\arrayrulecolor{gray!40}\cline{2-13}\arrayrulecolor{black}
& \mbox{geth-graphql-2025} & 71  & . & . & \cmark & \cmark & \cmark & \cmark & \makecell[c]{Panic-reach.\\(\texttt{nilPanic})} & . & . & . \\
\arrayrulecolor{gray!40}\cline{2-13}\arrayrulecolor{black}
& \mbox{geth-tracers-2024} & 66  & . & . & \cmark & \cmark & \cmark & \cmark & \makecell[c]{Analyzer routine\\(LOAD check)} & . & . & . \\
\midrule
\multirow{4}{*}{\shortstack[l]{\textit{Integer}\\\textit{Overflow}}}
& \mbox{p224-elliptic-2021} & 3.3 & . & . & . & . & . & . & . & . & . & . \\
\arrayrulecolor{gray!40}\cline{2-13}\arrayrulecolor{black}
& \mbox{fasthttp-2020}      & 6.5 & . & . & . & . & . & . & . & . & . & . \\
\arrayrulecolor{gray!40}\cline{2-13}\arrayrulecolor{black}
& \mbox{tendermint-2018}    & 34  & . & . & . & . & . & . & . & . & . & . \\
\arrayrulecolor{gray!40}\cline{2-13}\arrayrulecolor{black}
& \mbox{evm-gascost-2017}   & 19  & . & . & . & . & . & \cmark & \makecell[c]{Analyzer routine\\(INT\_MULT check)} & . & . & . \\
\midrule
\multirow{3}{*}{\shortstack[l]{\textit{Index Out}\\\textit{of Bounds}}}
& \mbox{kube-sm-2025}       & 87  & . & . & . & \cmark & \cmark & . & . & . & . & . \\
\arrayrulecolor{gray!40}\cline{2-13}\arrayrulecolor{black}
& \mbox{coredns-2025}       & 110 & . & . & . & \cmark & \cmark & \cmark & \makecell[c]{Panic-reach.\\(\texttt{panicIndex})} & . & . & . \\
\arrayrulecolor{gray!40}\cline{2-13}\arrayrulecolor{black}
& \mbox{goprotobuf-2013}    & 3.8 & . & . & . & \cmark & \cmark & \cmark & \makecell[c]{Panic-reach.\\(\texttt{panicIndex})} & . & . & . \\
\bottomrule
\end{tabular}%
}
\end{table*}

At that point, the overflow checker widens both 64-bit symbolic operands to 128~bits and queries Z3: if the upper half of the product can be non-zero, the multiplication can silently wrap. Z3 confirms the overflow in 0.39\,s and returns a concrete witness. No other tool in our evaluation can detect this bug without relying on a manually written oracle (Table~\ref{tab:go-tools-comparison}).

\begin{table}[h]
\centering
\caption{Overlay activity across the 11 binaries. \emph{(a)}~side bug caught during overlay execution at the indicated depth; \emph{(b)}~depth~15 reached, then an AST scan + Z3 confirmed a panic site on the same path; \emph{(c)}~no side bug.}
\label{tab:overlay-side-bugs}
\scriptsize
\renewcommand{\arraystretch}{1.0}
\setlength{\tabcolsep}{3pt}
\begin{tabular}{@{} l l l c @{}}
\toprule
\textbf{Binary} & \textbf{Side bug} & \textbf{Opcode} & \textbf{Depth} \\
\midrule
\multicolumn{4}{@{}l}{\emph{(a) Caught during overlay execution}} \\
kubelet-2025       & Concrete nil deref          & LOAD    & 3 \\
geth-graphql-2025  & Concrete nil write          & STORE   & 2 \\
\midrule
\multicolumn{4}{@{}l}{\emph{(b) Depth limit reached, confirmed on same path}} \\
kubelet-2025       & Reachable panic             & CBRANCH & 15 \\
geth-tracers-2024  & Reachable panicIndex        & CBRANCH & 15 \\
kube-sm-2025       & Reachable panic             & CBRANCH & 15 \\
goprotobuf-2013    & OOB slice access            & LOAD    & 15 \\
tendermint-2018    & Nil pointer deref           & LOAD    & 15 \\
tendermint-2018    & Reachable panic ($\times$3) & CBRANCH & 15 \\
\midrule
\multicolumn{4}{@{}l}{\emph{(c) No side bug:} kubectl-2025, coredns-2025, evm-gascost-2017,}\\
\multicolumn{4}{@{}l}{\quad fasthttp-2020, p224-elliptic-2021} \\
\bottomrule
\end{tabular}
\end{table}

\begin{table}[h]
\centering
\caption{Average detection time per tool, based on the 8 Go vulnerability cases found.}
\label{tab:detection-time}
\renewcommand{\arraystretch}{1.0}
\setlength{\tabcolsep}{6pt}
\small
\begin{tabular}{@{} l c @{}}
\toprule
\textbf{Tool} & \textbf{Avg.\ Detection Time} \\
\midrule
nilaway       & ${\approx}$2\,s \\
go test -fuzz & 0.18\,s \\
GoLibAFL      & ${\approx}$7\,s \\
Zorya         & 16.5\,min \\
\bottomrule
\end{tabular}
\end{table}

\section{Implementation}
\label{sec:implementation}

\subsection{Compiler-Aware Exploration}

Zorya adapts its vulnerability detection strategy based on the compiler. TinyGo inserts explicit calls to \texttt{runtime.nilpanic()} at nil checks, so AST exploration detects these panics. Overlay analysis is also enabled for TinyGo binaries because they can contain undefined behavior that do not trigger explicit panic calls. The \texttt{gc} compiler uses CPU traps for implicit nil dereferences (signal-based panics) \emph{and} explicit panic calls; both detection methods are therefore required~\cite{ralfs_ramblings_there_2025}. Since Zorya operates on Ghidra's P-Code, a language-agnostic intermediate
representation, it can also analyze C and C++ binaries. These binaries lack Go's panic infrastructure and rely solely on overlay execution for vulnerability detection.

\subsection{Multi-Thread State Management}

Zorya manages all OS threads present in the target binary. At initia\-lization, a \texttt{gdb} script dumps the register state, thread-local storage bases, and backtrace of every thread. Each thread is classified by its backtrace: the thread executing \texttt{main.main} is marked as the main thread, the \texttt{runtime.sysmon} thread is marked as the system monitor, and all others are marked as waiting.

The scheduler supports two policies. In \emph{main-only} mode, only the main thread executes. In \emph{round-robin} mode~\cite{rasmussen_round_2008}, threads switch cooperatively at function call boundaries after a configurable number of P-Code instructions. Thread switches occur only at function calls, mirroring Go's cooperative preemption model.

\subsection{Correctness, Soundness and Completeness}
Zorya is \emph{sound for the concrete path}: every instruction is executed with its real concrete value, and symbolic expressions mirror the concrete operations. On overlay paths, Zorya is \emph{best-effort} and may miss vulnerabilities beyond the depth limit or produce false positives when under-constrained symbolic variables yield satisfying but practically unreachable assignments. The tool is \emph{incomplete} by design: it follows a single concrete trace and explores untaken branches to a bounded depth, since complete path exploration is infeasible for \texttt{gc} binaries whose runtime alone contains hundreds of thousands of basic blocks.

\section{Evaluation}
\label{sec:eval}

We evaluate Zorya on real-world Go binaries compiled with \texttt{gc}, measuring detection accuracy, detection scope, and comparison with existing tools, under the following research questions:

\begin{itemize}
    \item \textbf{RQ1}: Can Zorya detect vulnerabilities in real-world multi-threaded binaries compiled with the Go compiler?
    \item \textbf{RQ2}: Can Zorya detect vulnerabilities not related to runtime panics?
    \item \textbf{RQ3}: How does Zorya compare to other Go tools and symbolic execution tools regarding vulnerability detection?
\end{itemize}

\subsection{Experimental Setup}
Experiments ran on 64-bit Linux with an Intel Core i9-12900K (24~threads, 3.2\,GHz) and 125\,GiB of RAM, using Zorya~v0.0.5 and Ghidra~v12. To ensure an accurate reproduction of the bugs, each target was compiled using the specific Go toolchain version that was current at the time its respective fix commit was reported. We evaluated Zorya against \texttt{staticcheck}, \texttt{gosec}, \texttt{nilaway}, \texttt{go~test~-fuzz}, \texttt{GoLibAFL}, \texttt{BINSEC}~v0.10.1, and \texttt{SymQEMU}. All tools were used with default configurations; where no versioned release was available, we used the latest code from the respective repositories as of January~2026. 

\subsection{Benchmark}
Our benchmark includes 11 real-world vulnerabilities from production Go projects, inspired by the Logic Bomb~\cite{xu_concolic_2017} approach of focused programs isolating specific runtime failures. Each vulnerability was derived from an actual bug fix: we identified the root cause from the commit message and patch, then compiled the vulnerable version of the project. For standalone applications we used the full production binary; for libraries, we compiled an example program from the project's official repository that exercises the faulty code path. All binaries are multi-threaded \texttt{gc} compilations. 

The vulnerabilities span three classes: \textbf{(i) Nil pointer dereferences, 4 cases:} uninitialized struct fields or unchecked pointers dereferenced before validation, from Kubernetes and Go-Ethereum. \textbf{(ii) Integer overflows, 4 cases:} silent arithmetic wrapping in unsigned or signed multiplication and accumulation, from Go-Ethereum, Tendermint, fasthttp, and Go's standard library. \textbf{(iii) Index out-of-bounds, 3 cases:} empty slices accessed without length checks or overflow-induced negative indices, from kube-state-metrics, CoreDNS, and golang/protobuf.

\subsection{Results and Analysis}
Table~\ref{tab:go-tools-comparison} summarizes the detection results across all 11 vulnerabilities and eight tools.

\textbf{RQ1: Multi-threaded \texttt{gc} binary analysis.}
Zorya correctly dumps the register state, TLS bases, and backtraces of all OS threads spawned by the \texttt{gc} runtime. It retrieves the runtime's internal offsets (goroutine descriptor, stack guard, processor struct) and neutralizes preemption before entering the target function. All 11 runs used the \texttt{main-only} scheduling policy, which restricts concolic execution to the main thread. This strategy is well suited to function-mode analysis: it enables continuous exploration of the function body without interruption from the garbage collector, the system monitor, or other runtime activities. Under this configuration, Zorya detects 7 of 11 vulnerabilities: four nil-pointer dereferences, one integer overflow, and two index out-of-bounds panics.

\smallskip
\noindent
\colorbox{gray!10}{%
    \parbox{0.47\textwidth}{
        \textbf{Finding 1:} Zorya successfully analyzes \texttt{gc}-compiled multi-threaded Go binaries and detects 7 out of 11 real-world vulnerabilities using function-mode analysis with main-only scheduling.
    }
}
\smallskip

\textbf{RQ2: Detection beyond runtime panics.}
Of the 7 primary detections, 4 are caught by the Analyzer Routine on the taken path, including the silent \texttt{INT\_MULT} overflow that no other tool finds, and 3 by AST panic\nobreakdash-reachability on the not\nobreakdash-taken branch.
Overlay execution further surfaces 10 side findings across 6 binaries (Table~\ref{tab:overlay-side-bugs}): 2 caught directly within the 15\nobreakdash-instruction depth exploration (kubelet at depth~3; geth\nobreakdash-graphql at depth~2), and 8 via an AST fallback once the exploration depth is reached, suggesting it should be increased. All correspond to functions that accept \texttt{nil} pointers or nil\nobreakdash-valued interfaces, a pattern permitted by Go but confirmed satisfiable and reachable by Z3.
The 5 remaining binaries yield no side finding: p224\nobreakdash-elliptic and fasthttp halt on heavy runtime helpers; the others reach depth~15 without exposing a new pattern.

\smallskip
\noindent
\colorbox{gray!10}{%
    \parbox{0.47\textwidth}{
        \textbf{Finding 2:} The 7 primary bugs are caught by the Analyzer Routine on the taken path (including the silent overflow unique to Zorya) or by the AST panic\nobreakdash-reachability. Overlay execution additionally surfaces side issues in 2 binaries.
    }
}
\smallskip

\textbf{RQ3: Comparison with other tools.}
Table~\ref{tab:go-tools-comparison} shows that static analyzers (\texttt{staticcheck}, \texttt{gosec}) detect none of the 11 bugs. \texttt{gosec}'s G115 rule flags type-conversion overflows (e.g., \texttt{uint64} to~\texttt{int}) but misses same-type arithmetic overflows. \texttt{nilaway} finds two nil-pointer dereferences in Go-Ethereum but cannot analyze Kubernetes packages due to dependency resolution failures. BINSEC and SymQEMU detect none of the 11 bugs; both lack support for Go's runtime primitives.

Fuzzers (\texttt{go~test~-fuzz} and \texttt{GoLibAFL}) are strong at detecting crash-producing bugs but require a harness for each tested function. They also detect \texttt{kube-sm-2025}, which Zorya misses: the vulnerable path traverses \texttt{runtime.memmove}, whose symbolic execution exhausts resources before reaching the fault site. For found bugs, fuzzers average a few seconds (Table~\ref{tab:detection-time}), whereas Zorya averages 16.5~minutes because it executes every instruction symbolically rather than only exploring the AST for known bug patterns, but in return produces an exact instruction\nobreakdash-level execution trace that fuzzers do not provide.

\smallskip
\noindent
\colorbox{gray!10}{%
    \parbox{0.47\textwidth}{
        \textbf{Finding 3:} Function-mode analysis is essential for complex real-world binaries: Zorya is slower than fuzzers but yields a full instruction-level trace and detects silent vulnerabilities without oracles or harnesses.
    }
}

\section{Discussion, Limits and Improvements}
\label{sec:limits}
\textbf{Symbolic analysis.}
Reaching a deep bug requires executing every preceding instruction. \emph{Symbolic summaries} for heavy runtime helpers and lightweight \emph{checkpointing} are the two natural mitigations. Extending symbolization to local variables would also broaden the detectable surface.

\textbf{Overlay Concolic Execution.}
The exploration depth, fixed to 15 instructions, warrants a more principled calibration. This approach is particularly suited to paths gated by nested conditionals.

\textbf{Evaluation dataset and concurrency.}
The evaluation binaries belong to the cloud and blockchain ecosystems and have over 5,000 GitHub stars. Issues were selected to cover vulnerability classes within Zorya's known detection scope. The round-robin scheduler has not yet been evaluated against concurrency-bug classes. Additional vulnerability patterns (e.g., division by zero, use-after-free) can be supported by extending the set of inspected P-Code operations. A more extensive evaluation on C corpora is planned~\cite{gorna_concolic_2026}.

\textbf{Other tools.}
BINSEC requires symbolic support for additional syscalls and CPU instructions to handle the Go runtime. SymQEMU would benefit from support for Go's cooperative preemption and multi-threaded initialization. Tools such as KLEE or SymCC~\cite{poeplau2020symbolic} require a functional Go-to-LLVM-IR compiler front-end. This study is limited to deterministic tools. It would be relevant to evaluate the corpus with machine-learning models for bug detection, but the approach would be fundamentally different.

\section{Conclusion}
\label{sec:conclusion}

This paper extends Zorya to binaries compiled with the \texttt{gc} compiler. Of the 7 primary detections, 4 are caught by the Analyzer Routine on the taken paths and 3 by AST panic\nobreakdash-reachability; overlay execution additionally surfaces 2 side findings on the not taken paths. The evaluation highlights a gap in the Go security landscape: static analyzers miss bugs requiring program\nobreakdash-specific arithmetic or aliasing reasoning, fuzzers require per\nobreakdash-function harnesses, and binary\nobreakdash-level symbolic executors cannot yet handle the Go runtime. Zorya is slower than fuzzers but fully automatic and able to detect silent vulnerabilities that produce no crash signal. Future work targets broader symbolic coverage and concurrency\nobreakdash-bug detection.

\paragraph*{\textbf{Acknowledgment.}} The authors thank the anonymous reviewers for their valuable feedback and the Ledger Donjon, Telecom Paris, and University of the Western Cape teams for their support.

\bibliographystyle{ACM-Reference-Format}
\bibliography{sample-base}

\end{document}